\documentclass[10pt,conference]{IEEEtran}
\IEEEoverridecommandlockouts

\usepackage{cite}
\usepackage{amsmath,amssymb,amsfonts}
\usepackage{algorithmic}
\usepackage{graphicx}
\usepackage{textcomp}
\usepackage{xcolor}
\usepackage{enumitem}
\usepackage{microtype}
\usepackage{url}

\def\BibTeX{{\rm B\kern-.05em{\sc i\kern-.025em b}\kern-.08em
    T\kern-.1667em\lower.7ex\hbox{E}\kern-.125emX}}

\IEEEoverridecommandlockouts
\IEEEpubid{\makebox[\columnwidth]{978-3-903176-15-7~\copyright~2019 IFIP \hfill}
\hspace{\columnsep}\makebox[\columnwidth]{ }}

\begin{document}

\title{Multi-party authorization and conflict mediation for decentralized configuration management processes
\thanks{This work has been supported by the German Federal Ministry of Education
and Research, project DecADe, grant 16KIS0538 and the German-French
Academy for the Industry of the Future. 
\protect\\
\indent Author's version -- Final paper to appear in 2019 IEEE/IFIP International Workshop on Hot Topics in Network and Service Management (HotMSM) co-located with the  IFIP/IEEE International Symposium on Integrated Network Management (IM).}
}

\author{\IEEEauthorblockN{Holger Kinkelin, Heiko Niedermayer, Marc Müller and Georg Carle}
\IEEEauthorblockA{Technische Universität München, Department of Informatics,
Chair of Network Architectures and Services \\
85748 Garching bei München, Germany\\
\{lastname\}@net.in.tum.de}}

\maketitle

\begin{abstract}

Configuration management in networks with highest security demands must not depend on just \emph{one} administrator and her device.
Otherwise, problems can be caused by mistakes or malicious behavior of this admin, or when her computer got compromised, which allows an attacker to abuse the administrator's far-reaching permissions.

Instead, we propose to use a reliable and resilient configuration management process orchestrated by a \emph{configuration management system (CMS)}.
This can be achieved by
\emph{separation of concerns} (proposing a configuration vs. authorizing it),
employing \emph{multi-party authorization (MPA)},
and enforcing that only authorized configurations can be deployed.
This results in a configuration management process that is \emph{decentralized} on a human, decision-making level, and a technical, device level.

However, due to different opinions or adversarial interference, the result of an MPA process can end in a conflict.
This raises the question how such conflicts can be mediated in a better way than just employing majority voting, which is insufficient in certain situations.
As an alternative, this paper introduces building blocks of customizable conflict mediation strategies which we integrated into our CMS \emph{TANCS}~\cite{8406324}.
The conflict mediation functionality as well as the initial TANCS implementation run on top of the distributed ledger and smart contract framework Hyperledger Fabric which makes all processes resilient and tamper-resistant.

\end{abstract}


\section{Introduction}
\label{sec:intro}

In the age of advanced persistent threats (APT), it is almost safe to predict that any corporate, governmental, or military network that is important enough will be compromised sooner or later.
The effects of such attacks range from data leaks to tampered control processes of critical infrastructures.

One crucial factor for network security is meticulous configuration management.
However, management processes that depend on just \emph{one} administrator and her device can fail and cause problems.

An inexperienced or careless administrator, for instance, might accidentally  deploy a faulty configuration to managed devices and cause an availability problem, or create a security weakness that can be exploited later.

Even more problematic are disgruntled insiders who actively try to harm the organization they are working for~\cite[p.~53]{istr16}~\cite[p.~10]{beyond}.
For this reason, an admin who turns malicious is a major threat.

Besides internal issues, problems related to network management also emerge from the outside of the network.
In APTs, attackers often target administrators directly using water holing or spear phishing attacks~\cite[p.~28]{istr18} pursuing the target of compromising the admin's account or computer.

This is highly problematic, as admins often use just one cryptographic key to authenticate themselves to managed entities.
If the attacker has access to this key, she is able to manage the network as she pleases.

In the remaining paper, we argue for decentralizing configuration management in order to avoid single points of failure on technical and in particular on human level.
For this purpose, we detail the benefits of \emph{separation of concerns} in management processes and of \emph{multi-party authorization (MPA)} of new configurations before deploying them, see Sect.~\ref{sec:avoidspof}.
Due to different opinions and also adversarial interference, MPA processes of new configurations can end in a conflict.
To better understand the causes of such error cases, we conduct an analysis in Sect.~\ref{sec:causes}.
Based on these insights, we argue why a simple majority vote is not enough to mediate such conflicts.
As an alternative we present building blocks which can be assembled to customizable conflict mediation strategies, see Sect.~\ref{sec:blocks}.
The next Sect.~\ref{sec:implout} outlines how we integrated discussed building blocks and mediation strategies with our previously published configuration management system \emph{TANCS}.
Lastly, we discuss our work in Sect.~\ref{sec:discussion}, compare it to related work, c.f. Sect.~\ref{sec:prwork}, and conclude the paper in Sect.~\ref{sec:conclusion}.

\section{Avoiding single points of failure by decentralized configuration management}
\label{sec:avoidspof}

As we already motivated,
one careless administrator who accidentally deploys a faulty configuration,
one administrator who intentionally deploys a malicious configuration,
or one compromised device or cryptographic key abused to deploy a malicious configuration
can be enough to pose a serious threat to the security and availability of a networked IT system.

For this reason, we argue that it is time to rethink configuration management processes from a reliability and resilience standpoint, which suggests that single points of failure should be avoided.

Instead of having a \emph{centralized} configuration management process where only one admin can create and immediately deploy a new configuration, we propose to split the configuration management process in different steps.
These steps all need to be performed by different persons or entities using different devices, and are orchestrated by an entity called \emph{configuration management system (CMS)}.

In the \emph{proposal phase}, an administrator \emph{proposes} a new configuration and hands it over to the CMS.
In the \emph{review phase}, several reviewers independently assess the configuration and individually communicate their approval or refusal to the CMS.
In the \emph{authorization phase}, the CMS decides whether to finally accept and deploy, or to refuse the configuration based on the input of reviewers.

The result of the just described separation of concerns combined with the multi-party authorization (MPA) process of the proposed configuration is \emph{decentralized}, which is beneficial on two levels:
First, the distribution of the configuration management process to various \emph{persons} prevents on a human, decision-making level that an individual careless or malicious administrator can cause harm to the network.
Second, the distribution of the process to several \emph{devices} used by human actors prevents on a technical, device level that individual compromised devices or cryptographic keys can be successfully abused to deploy a malicious configuration.

\section{Analysis of conflicts and attacks on configuration management processes}
\label{sec:causes}

This section analyzes causes of conflicts in MPA processes with regard to decentralized configuration management.
For our analysis we use the different process phases as described in Sect.~\ref{sec:avoidspof}.
We however assume that the authorization phase cannot be tampered with easily when a tamper-resistant CMS like TANCS, see Sect.~\ref{sec:tancs}, is used.
For this reason, we focus on the proposal and review phase, which are close to human actors, and hence prone to faults and manipulation.

\paragraph{Proposal phase}

In this phase,

a proposer can either propose a \emph{valid} or \emph{invalid} configuration.
Valid configurations are correct and suitable for a managed device.
Invalid configurations are either faulty when proposed from a careless administrator, or malicious when proposed from a rogue admin or an attacker from remote.

\paragraph{Review phase}

In this phase,

each reviewer can either agree to or reject a proposed configuration.
The result of this step can either correspond to or conflict with the actual validity of the proposed configuration.
We consider agreeing to an invalid configuration as either a flaw or an attack.
Similarly, rejecting a valid configuration is either a flaw or an attack.

\section{Building blocks for mediation strategies of MPA conflicts}
\label{sec:blocks}

Sect.~\ref{sec:avoidspof} discussed the advantages of decentralizing a configuration management process by separation of concerns and multi-party authorization (MPA).
Due to the reasons given in Sect.~\ref{sec:causes}, the MPA process can end in a conflict.
Therefore, a mechanism that mediates such conflicts is needed.

In situations where human beings have different opinions about a subject, consensus is traditionally achieved by mechanisms like majority voting.
Majority voting, however, returns the wrong result when the minority of answers is right and the majority is wrong.
Such a scenario emerges, for instance, if one careful reviewer finds a flaw or a malicious statement in a configuration that other reviewers did not spot.
With majority voting, the configuration would be approved which invalidates the benefits of decentralizing the decision-making process.

However, majority voting would be sufficient in other scenarios.
In cases where one adversary has compromised a reviewer's machine and tries to stop a valid configuration by refusing it, overruling the minority would yield the right decision, i.e. accepting this configuration.

This dilemma shows that more elaborate and targeted conflict mediation strategies than majority voting are needed in networks with very high security demands.
The two goals of a suitable conflict mediation strategy must be
1) to increase the chance of rejecting a malicious configuration, and
2) to increase the chance of accepting a valid configuration when under attack.
For this purpose, we propose various building blocks (BB) that can be assembled to different conflict mediation strategies as outlined in Sect.~\ref{sec:combiningblocks}.

The essential idea of this approach is to perform additional ``rounds'' after the initial MPA yielded a conflict. In these rounds, reviewers are enabled to rethink and change their opinion, and to commit a new decision. At the end of a round, the conflict is either resolved -- as the CMS finally accepts or refuses the configuration -- or a further round needs to be executed.

\subsection{BB1: Request confirmation}

BB1 pursues the idea of informing reviewers that there is a conflict, and giving them the opportunity to correct their answer.
As additional information, each reviewer receives short ``commit messages'' entered by other reviewers that explain why they, for instance, rejected the respective configuration in a previous round.

BB1 is helpful in situations where individual reviewers did not notice a problem in a configuration.
Using the additional input, they most likely spot the problem and correct their decision, which helps to reach consensus on this configuration.

\subsection{BB2: Confirmation via 2nd channel}

If we consider adversaries able to control devices from remote, it is also possible that a reviewer got impersonated and did not even notice that, for instance, a valid configuration was rejected from her computer.
Further interaction with this reviewer is not possible as the adversary can intercept and answer every inquiry from the CMS.

For this reason, BB2 tries to evade the compromised device by using a second -- hopefully still trustworthy -- channel to the reviewer, e.g. using a second device.
Using this channel, a reviewer can be requested to confirm that a particular review decision committed from her computer to the CMS was indeed given by her.
If the admin does not confirm, the CMS can exclude the compromised device from the process.

\subsection{BB3: Incorporate additional reviewers}

BB3 follows the approach to incorporate additional reviewers in the process. Situations where this is beneficial include replacing reviewers whose computer got compromised.

BB3 is also helpful to collect additional information from new reviewers, which can be used in BB1 to allow reviewers to rethink their decision.

\subsection{BB4: Direct conflict mediation via chat}

We regard BB1, 2 and 3 as still being only modestly interrupting and hence being quite ``inexpensive''.
However, we expect that it is not always possible to achieve consensus using these building blocks.
For this reason, BB4 and 5 follow the idea to enable reviewers to resolve a conflict in a more direct and interactive manner.

BB4 adds a chat-like function to the CMS, which enables a direct discourse between reviewers.
To avoid that a single reviewer is able to give the final decision of the whole group, all involved reviewers individually commit the group's decision to the CMS.

\subsection{BB5: Direct conflict mediation in person}

BB5 is a variant of BB4. However, instead of trying to mediate the conflict in a chat, reviewers must meet and mediate the situation in person.
As in BB4, each reviewer commits individually the agreement on decision to the CMS to prevent that a single reviewer can give this final decision.
An additional benefit of BB5 is that it helps when all communication channels to a reviewer are compromised.

\subsection{Examples of composite conflict mediation strategies}
\label{sec:combiningblocks}

As we have discussed, each of the building blocks has different properties and associated costs as it requires more or less additional effort from reviewers.
For this reason, conflict mediation strategies can be tailored to different security requirements of managed devices or situations.
These strategies can be defined on a per device basis or for groups of devices with similar security requirements, and be worked off by the CMS when the initial MPA process ends in a conflict.

Mediating a conflict that concerns a group of highly important entities, like the network's firewall or identity management system, is worth investing a lot of effort.
So, a mediation strategy that includes all building blocks, maybe even repeatedly, could be specified.

Investing the same effort is maybe inappropriate for a group of lesser relevant components in the network.
In such cases, the mediation strategy does not include expensive building blocks like BB3 - BB5.
Instead, the system will abort conflict mediation and reject the configuration after BB2 is finished without consensus.

\section{Implementation}
\label{sec:implout}

We added the described conflict mediation functionality to our CMS \emph{TANCS}~\cite{8406324}.
TANCS stands for \textbf{t}amper-resistant and \textbf{a}uditable \textbf{n}etwork \textbf{c}onfiguration management \textbf{s}ystem.

\subsection{TANCS}
\label{sec:tancs}

TANCS is able to conduct and enforce the configuration management process described in Sect.~\ref{sec:avoidspof}, i.e., it requires that multiple human experts review and approve a new configuration.
Only if a configuration has been accepted by \emph{all} reviewers specified in a device-specific policy, the current TANCS implementation will set the status of this configuration to \emph{authorized}.
Managed devices, which are required to be locked down to prevent other configuration mechanisms, automatically pull authorized configurations addressed to themselves from TANCS and apply them locally.
Besides this functionality, we ensure accountability and traceability of the entire process for forensic purposes.

TANCS runs on top of Fabric~\cite{hlfos}, which is a distributed ledger and smart contract framework developed by the Hyperledger project of the Linux Foundation.
Every input from an administrator or reviewer sent via a command line client (CLI) to TANCS is processed by a smart contract running on multiple nodes in the Fabric peer-to-peer network.
Furthermore, all in- and outputs of these operations are stored in a redundant, non-modifiable and inerasable manner in the distributed ledger established by the Fabric peers.
As long as the majority of nodes is honest, individual adversaries are not able to forge or erase the outcomes of the configuration management process, which is why TANCS is tamper-resistant.

A further interesting fact is that TANCS is inherently able to support configuration management of IT infrastructure shared across different stakeholders, who are not even required to trust each other.
In such cases, every stakeholder participates with reviewers and Fabric peers, which both represent this stakeholder's interests.

\subsection{Conflict mediation functionality}

The described conflict mediation building blocks and the concept of combining them to different strategies were implemented and added to TANCS using the same paradigms as used for the initial TANCS functionality:

Individual building blocks were implemented as new smart contracts. They interact with reviewers via the CLI, and process and persist the input of reviewers in suitable data structures, which are stored in the distributed ledger.

Different conflict mediation strategies can be expressed as policies that are applied to devices or device groups.
Each policy refers to those building blocks that shall be part of the respective conflict mediation strategy.

In the case of MPA yielding a conflict, or after a round could not mediate the conflict, TANCS determines the next building block as defined by the strategy and executes it.
The evaluation of the result after each mediation round is likewise administered by a policy-evaluation smart contract executed on the peers. The state of the entire process is persisted in the distributed ledger.

\section{Discussion}
\label{sec:discussion}

\subsection{Costs vs. benefits of using a CMS}

Compared to simply logging in and running a command to configure a device, the overhead of using a CMS with MPA and conflict mediation seems to be huge.
However, the CMS helps to prevent that invalid configurations can be deployed.
So, operating a CMS is most likely less expensive than recovering from a severe outage, or losing customers due to a serious data leak, etc.
The additional conflict mediation strategies proposed in this paper add further cost to the CMS.
Conflict mediation should only happen occasionally and resulting costs are well invested as the conflict needs to be dealt with anyway.

\subsection{Conflict mediation vs. majority voting}

The next question is whether conflict mediation increases the chance
1) to reject invalid configurations and
2) to accept valid configurations when under attack
compared to majority voting.
For this discussion, we

assume that most of reviewers and devices are still benign.

When we assume that an invalid configuration got proposed, MPA increases the chance that at least one reviewer will spot the error in the first round.
As a result, this configuration is stopped and the resulting conflict must be mediated in subsequent rounds where we enable reviewers to reconsider their review by pointing them to the problem.
Honest reviewers that did not spot the problem and accepted the configuration, will most likely notice the problem and reject the configuration.
This helps to reach consensus and to reject the invalid configuration for good.
Vice versa, adversarial reviewers that accepted the invalid configuration can be identified when they keep accepting the invalid configuration in subsequent rounds.
This helps to exclude adversaries.

When we assume that a valid configuration got proposed, MPA increases the chance that it is not rejected for good in the first round.
In subsequent rounds, honest reviewers that accidentally rejected the configuration, can be convinced by others that the configuration is valid and to finally accept it.
Vice versa, adversarial reviewers that repeatedly try to convince others to reject the valid configuration can be identified.

Additionally, BB2 (2nd channel) can actively unveil compromised hosts, which helps in both conflict cases to identify adversaries.

\subsection{Open issues}

MPA and conflict mediation creates a lot of delay which can be problematic in emergency situations where quick responses are needed.
In emergency situations, the CMS might allow an administrator to override the MPA process and to directly deploy a configuration.
This, however, creates a loophole for possibly malicious admins which cannot be avoided if such a feature is required.
One way of mitigating the situation is based on logging which adds accountability of all actions performed by the admin while overriding the MPA process.

\section{Related Work}
\label{sec:prwork}

\paragraph{MPA in open-source projects}

Recently, the problem of infecting the software supply chains got more attention~\cite[p. 42]{istr18}.
This includes injecting malicious code in open-source projects hosted on services like GitHub or GitLab.
This is possible as such projects typically allow unknown contributors to propose code changes, which then need to be accepted and integrated by the project's maintainers.

Because of this and other reasons, MPA starts to be supported by GitHub~\cite{github} and GitLab~\cite{gitlab}.
While GitHub only allows to specify the number of reviewers, GitLab additionally allows to specify which maintainers must perform a code review before the change is added to the code base.
As a difference to our approach, GitHub and GitLab use processes running on centralized servers that control the MPA process. Our solution is based on distributed ledger and smart contract technology.

\paragraph{Distributed consensus and fault tolerance}

Distributed consensus and fault tolerance problems deal with maintaining the correct current state among good peers as long as the malicious ones are a small enough minority.
Such -- partially Byzantine fault tolerant -- protocols have been extensively studied~\cite{10.1007/3-540-12689-9_99, Dfago2000TotallyOB, Castro:1999:PBF:296806.296824}.
While we use such algorithms as part of the distributed ledger, they do not fit directly to the human decision-making process in a CMS as they do not factor in human knowledge.

\section{Conclusions}
\label{sec:conclusion}

In this paper, we pointed out that centralized configuration management processes must be avoided.
Instead, we proposed to use a reliable and resilient, decentralized process controlled by a configuration management system (CMS).
Such a process can be created by the means of separation of concerns and multi-party authorization (MPA).

However, as MPA can result in conflicts, we proposed configurable conflict mediation strategies that pursue two goals, namely increasing the chance
1) to reject malicious configurations and
2) to accept good configurations when under attack.
We discussed the benefits of our approach over majority voting and finally described how strategies can be implemented as part of our tamper-resistant and auditable configuration management system TANCS.

Future work includes a more formal analysis of TANCS and an extension of the life cycle of configurations that allows to quickly switch between authorized configurations in urgent situations like network failures or attacks.

\IEEEtriggeratref{0}
\bibliographystyle{IEEEtran}
\bibliography{ref}

\begin{thebibliography}{10}
\providecommand{\url}[1]{#1}
\csname url@samestyle\endcsname
\providecommand{\newblock}{\relax}
\providecommand{\bibinfo}[2]{#2}
\providecommand{\BIBentrySTDinterwordspacing}{\spaceskip=0pt\relax}
\providecommand{\BIBentryALTinterwordstretchfactor}{4}
\providecommand{\BIBentryALTinterwordspacing}{\spaceskip=\fontdimen2\font plus
\BIBentryALTinterwordstretchfactor\fontdimen3\font minus
  \fontdimen4\font\relax}
\providecommand{\BIBforeignlanguage}[2]{{%
\expandafter\ifx\csname l@#1\endcsname\relax
\typeout{** WARNING: IEEEtran.bst: No hyphenation pattern has been}%
\typeout{** loaded for the language `#1'. Using the pattern for}%
\typeout{** the default language instead.}%
\else
\language=\csname l@#1\endcsname
\fi
#2}}
\providecommand{\BIBdecl}{\relax}
\BIBdecl

\bibitem{8406324}
H.~Kinkelin \emph{et~al.}, ``Trustworthy configuration management for networked
  devices using distributed ledgers,'' in \emph{NOMS 2018 - 2018 IEEE/IFIP
  Network Operations and Management Symposium}, April 2018.

\bibitem{istr16}
{Symantec Corporation}, ``{Internet Security Threat Report, Vol. 21},'' 2016.

\bibitem{beyond}
{Google Inc.}, ``{Google Infrastructure Security Design Overview},'' 2017,
  [Online] \url{https://cloud.google.com/security/security-design/}, last
  access \today.

\bibitem{istr18}
{Symantec Corporation}, ``{Internet Security Threat Report, Vol. 23},'' 2018.

\bibitem{hlfos}
E.~Androulaki \emph{et~al.}, ``Hyperledger fabric: A distributed operating
  system for permissioned blockchains,'' in \emph{Proceedings of the Thirteenth
  EuroSys Conference}.\hskip 1em plus 0.5em minus 0.4em\relax New York, NY,
  USA: ACM, 2018.

\bibitem{github}
{Bryan Clark}, ``{Require multiple reviewers for pull requests},'' 2018,
  [Online]
  \url{https://blog.github.com/2018-03-23-require-multiple-reviewers/}, last
  access \today.

\bibitem{gitlab}
{GitLab Inc.}, ``{Merge request approvals},'' 2018, [Online]
  \url{https://docs.gitlab.com/ee/user/project/merge_requests/merge_request_approvals.html/},
  last access \today.

\bibitem{10.1007/3-540-12689-9_99}
M.~J. Fischer, ``The consensus problem in unreliable distributed systems (a
  brief survey),'' in \emph{Foundations of Computation Theory}, M.~Karpinski,
  Ed.\hskip 1em plus 0.5em minus 0.4em\relax Berlin, Heidelberg: Springer
  Berlin Heidelberg, 1983, pp. 127--140.

\bibitem{Dfago2000TotallyOB}
X.~D{\'e}fago and A.~Schiper, ``Totally ordered broadcast and multicast
  algorithms : A comprehensive survey,'' 2000.

\bibitem{Castro:1999:PBF:296806.296824}
M.~Castro and B.~Liskov, ``Practical byzantine fault tolerance,'' in
  \emph{Proceedings of the Third Symposium on Operating Systems Design and
  Implementation}.\hskip 1em plus 0.5em minus 0.4em\relax Berkeley, CA, USA:
  USENIX Association, 1999.

\end{thebibliography}

\end{document}